\documentclass[notitlepage,english,aps,floats,onecolumn,showpacs,nofootinbib,floatfix]{revtex4-1}

\usepackage{pslatex}
\usepackage[T1]{fontenc}
\usepackage[latin1]{inputenc}
\usepackage{graphicx}
\usepackage{epsfig}
\usepackage{longtable}
\usepackage{float}
\usepackage{calc}
\usepackage{ifthen}
\usepackage{amsmath}
\usepackage{hyperref}
\usepackage{amssymb}

\usepackage{color}

{
{
{
\newcommand{\bea}{\begin{eqnarray}}
\newcommand{\eea}{\end{eqnarray}}

\newcommand{\nc}{\newcommand}
\nc{\renc}{\renewcommand}
\nc{\eqs}[2]{\mbox{Eqs.~(\ref{#1},\,\ref{#2})}}
\nc{\eq}[1]{\mbox{Eq.~(\ref{#1})}}
\nc{\figs}[2]{\mbox{Figs.~(\ref{#1},\,\ref{#2})}}
\nc{\fig}[1]{\mbox{Fig~.(\ref{#1})}}
\nc{\be}[1]{\begin{equation} \mbox{$\label{#1}$}}
\nc{\ee}{\vspace{0.1cm}\end{equation}}

\newcommand{\bean}{\begin{eqnarray*}}
\newcommand{\eean}{\end{eqnarray*}}

 \def\gae{\; ^{>}_{\sim} \;}

\begin{document}

\title{Does Palatini Higgs Inflation Conserve Unitarity?}
\author{J. McDonald }
\email{j.mcdonald@lancaster.ac.uk}
\affiliation{Dept. of Physics,  
Lancaster University, Lancaster LA1 4YB, UK}

\begin{abstract}

In the conventional metric formulation of gravity, the Higgs Inflation model violates unitarity in the electroweak vacuum in Higgs scattering at the energy scale $\Lambda \sim M_{Pl}/\xi$, where $\xi \sim 10^4$ is the non-minimal coupling of the Higgs to the Ricci scalar. In the Palatini formulation it is commonly believed that $\Lambda \sim M_{Pl}/\sqrt{\xi}$, where $\xi \sim 10^{9}$. Here we reconsider unitarity violation in the electroweak vacuum in the Palatini formulation. We argue that there is no unitarity violation in Higgs scattering in the Palatini non-minimally coupled Standard Model in the electroweak vacuum at energies below the Planck scale. In this case Palatini Higgs Inflation completely conserves unitarity and is consistent at all energies up to those at which quantum gravity becomes important. If true, this would imply that Palatini Higgs Inflation has a significant advantage over metric Higgs Inflation.    

\end{abstract}
 
\maketitle

\section{Introduction}   Higgs Inflation \cite{bs} (for a recent review see \cite{rubio}) is an important application of the non-minimally coupled scalar inflation model originally proposed in \cite{bbs}. It amounts to identifying the scalar of the model with the Standard Model Higgs boson. The attraction of this class of model is that it can achieve inflation with scalar fields which have a large quartic self-coupling, allowing scalars in conventional particle physics theories to explain inflation. However, while the model works well classically and is in excellent agreement with Planck observations, it encounters a serious problem at the fundamental quantum level, since unitarity is violated in tree-level Higgs scattering at energies much less than the Planck scale. Since some modification of the theory must occur in order to ensure a unitary theory\footnote{In this we are assuming that tree-level unitarity violation implies true unitarity violation and not simply a breakdown of perturbation theory \cite{sh,uvjr2}.}, we cannot be sure that the classical theory is consistent at energies or field values larger than the unitarity violation scale. In general, new non-renormalisable terms consistent with all symmetries and scaled by the unitarity-violation scale would be expected in the potential. (For discussions of the unitarity problem of non-minimally coupled inflation models, see for example \cite{uvjr2, uvbg1, uvhz, uvjr1, he, uvfk}. The existence of a strong coupling scale of the order of $M_{Pl}/\xi$ was first noted in \cite{jb}.) 

The origin of the unitarity violation problem in Higgs Inflation is the existence of more than one non-minimally coupled scalar degree of freedom. In the case where there is only a single non-minimally coupled real scalar\footnote{The absence of unitarity violation in the case of a single scalar was first shown in \cite{uvjr0}.}, there is a cancellation of the tree-level graviton-exchange diagrams that would otherwise cause the unitarity problem in the Jordan frame (the s-, t- u-diagram cancellation) \cite{uvhz}. In the Einstein frame this is manifested as a redefinition of the scalar to a canonically normalised field \cite{uvbg1, uvjr0}. However, with an additional scalar, the Jordan frame cancellation is no longer possible, as the necessary s-channel diagram for $\phi_{1} \phi_{2} \rightarrow \phi_{1} \phi_{2}$ does not exist. Equivalently, in the Einstein frame, the Lagrangian kinetic terms cannot be redefined to a sum of canonically normalised kinetic terms \cite{uvbg1, uvhz}.

There is an attractive alternative to metric Higgs Inflation that retains the advantages and successful predictions of the model, namely Palatini Higgs Inflation \cite{pi1, pi2}. (For recent studies of Palatini Higgs inflation see \cite{rp1}. For a review see \cite{tenkanen}.) In this formulation, the connection is treated as an independent variable, and only assumes the Levi-Civita form once the equations of motion are solved. The Ricci tensor is therefore independent of the metric. This changes the conformal transformation to the Einstein frame and the resulting inflation theory. In this note, we reconsider the question of unitarity violation in the electroweak vacuum due to the non-minimal coupling in Palatini Higgs Inflation. Whilst our argument is primarily a physical rather than a formal proof, it strongly suggests that, in so far as Higgs Inflation can be quantised and tree-level scattering amplitudes can be calculated at all energies, unitarity is completely conserved in the Palatini model at all energies up to the Planck scale in the electroweak vacuum and therefore there is no unitarity problem\footnote{This applies to conventional Palatini inflation with a non-minimal inflaton coupling to $R$. An alternative class of Palatini inflation models, based on the inclusion of an $R^2$ term \cite{prs2,ajp}, has quite different properties with respect to unitarity violation \cite{ajp}.}.

It is important to distinguish the case where unitarity is completely conserved, meaning no unitarity violation in the electroweak vacuum at all energies, from the case where inflation is consistent with unitarity violation in the presence of a background inflaton field $\overline{\phi}$. In the case of Palatini Higgs Inflation, the scale of unitarity violation in the Einstein frame is in fact independent of the background Higgs field and is of the order of $M_{Pl}/\sqrt{\xi}$ (we will confirm this in the present analysis). Essentially the same result is obtained in the case of metric Higgs Inflation, where the cut-off is of the order of the gauge boson mass. This scale corresponds to a Jordan frame scale of the order of $\overline{\phi}$ in both the Palatini and metric formalisms.  The Einstein frame unitarity violation scale is large compared to the Hubble parameter in the Einstein frame during inflation, which is of the order of $M_{Pl}/\xi$ and  determines the energy of interactions during inflation. Therefore 
it is argued that Higgs Inflation is consistent with unitarity violation. However, even assuming that a unitary completion which varies with the Higgs field actually exists, which has never been demonstrated, we would expect non-renormalisable terms scaled by the Jordan frame unitarity violation scale to appear in the Jordan frame effective theory. This would result in large corrections to the potential, invalidating both the metric and Palatini models. We will argue in the following that Palatini Higgs Inflation, unlike metric Higgs Inflation, may evade this problem completely.

\section{The Palatini Higgs Inflation Model and Unitarity Violation}

When considering scattering in the electroweak vacuum at energies $E \gg m_{W}$, the Goldstone equivalence theorem
 \cite{et}  tells us that we can consider the scalars in the Higgs doublet $H$ to be physical and neglect the gauge fields when computing tree-level scattering amplitudes. We will therefore consider the theory with only the Higgs scalars and no gauge fields from this point on. 
It will be into this model that we will later introduce a background Higgs field in order to estimate the unitarity violation scale at high energy in the electroweak vacuum.  

 The action of the model is 
\be{e1} S = \int d^{4} x \sqrt{-g} \left[ \left( 1 + \frac{2 \xi  |H|^{2}}{M_{Pl}^{2}} \right) \frac{M_{Pl}^{2} R}{2}   - \partial_{\mu} H^{\dagger} \partial^{\mu} H  - V(|H|) \right]  ~.\ee
We do not consider the potential in this analysis, as we are interested specifically in unitarity violation purely due to the non-minimal coupling, which determines the conventional unitarity violation scale in Higgs Inflation.

To estimate the unitarity violation scale we will transform to the Einstein frame, where the calculation is generally easier since the scalars are minimally coupled to gravity. The underlying principle is that unitarity violation is a physical process which is independent of the variables used to describe the action. If it breaks down in one frame at some energy (as defined in that frame) then it must break down at an equivalent energy in a different frame - an action cannot be unitary in one frame but not unitary in another. This means that if there is a unitarity violating process in the Jordan frame, there will be an equivalent unitarity violating process in the Einstein frame.  We therefore perform a conformal transformation to the Einstein frame, $\tilde{g}_{\mu\;\nu}  = \Omega^{2} g_{\mu\;\nu}$, where
\be{e2} \Omega^{2} =  1 + \frac{2 \xi |H|^{2}}{M_{Pl}^{2}} ~.\ee 
We define the Higgs doublet scalars by $H = ( (\phi_{1} + i \phi_{2})/\sqrt{2}, (\phi_{3} + i \phi_{4})/\sqrt{2} )$. In order to analyse unitarity violation, it is only necessary to consider one pair of real scalars, as scattering involving other pairs of scalars will be equivalent since the action \eq{e1} is symmetric under the exchange of any pair of real scalars. Restricting to $\phi_{1}$ and $\phi_{2}$, the Einstein frame action is given by  
\be{e3}  S =  \int d^{4} x \sqrt{-g} \left[ {\cal L}_{11} + {\cal L}_{12} + {\cal L}_{22} \right] ~,\ee 
where 
\be{e4} {\cal L}_{11} = -\frac{1}{2} \left( \frac{\Omega^{2} + 
\frac{6 \alpha \xi^2 \phi_{1}^{2}}{M_{Pl}^{2}} }{\Omega^{4}} \right) \partial_{\mu} \phi_{1}  \partial^{\mu} \phi_{1}   ~\ee
\be{e5} {\cal L}_{22} = -\frac{1}{2} \left( \frac{\Omega^{2} + 
\frac{6 \alpha \xi^2 \phi_{2}^{2}}{M_{Pl}^{2}} }{\Omega^{4}} \right) \partial_{\mu} \phi_{2}  \partial^{\mu} \phi_{2}   ~\ee
\be{e6} {\cal L}_{12} = - \frac{6 \alpha \xi^{2} \phi_{1} \phi_{2} \partial_{\mu} \phi_{1}  \partial^{\mu} \phi_{2} }{\Omega^{4} M_{Pl}^{2} }      ~.\ee
Here $\alpha = 0$ for Palatini and 1 for metric. 

To determine the scale of unitarity violation in the electroweak vacuum, we can consider scattering in the Einstein frame \cite{uvjr0,uvbg1,uvjr1}. In general, unitarity violation in $2 \rightarrow 2$ elastic scattering occurs at the energy scale where the magnitude of the scattering amplitude satisfies $|{\cal M}| \sim 1$ \cite{uvjr1}. Dimensionally, this occurs at the mass scale of the non-renormalisable scattering interaction.  
In the case of metric Higgs Inflation, the ${\cal L}_{12}$ term unambiguously violates unitarity at an energy $\Lambda \sim M_{Pl}/\xi$, assuming that $\Omega = 1$ at the scattering energy. We say 'unambiguously' since at the energy $\Lambda$, if translated into an expectation value for the fields $\phi_{1}$ and $\phi_{2}$ (the significance of this will be discussed below), we have $\Omega^{2} = 1 + O(1/\xi) \approx 1$, so the $1/\Omega^{4}$ term in ${\cal L}_{12}$ will not modify this conclusion.  

In the case of Palatini Higgs Inflation, the ${\cal L}_{12}$ operator does not arise and the Einstein action becomes simply 
\be{e7} S =  \int d^{4} x \sqrt{-g} \left[ - \frac{1}{2 \Omega^{2}} \partial_{\mu} \phi_{1} \partial^{\mu} \phi_{1} 
 - \frac{1}{2 \Omega^{2}} \partial_{\mu} \phi_{2} \partial^{\mu} \phi_{2}  \right]     ~,\ee 
where 
\be{e8} \Omega^{2} = 1 + \frac{\xi \left(\phi_{1}^{2} + \phi_{2}^{2} \right) }{ M_{Pl}^{2}}    ~.\ee 
In this case, unitarity violation would be due to the $\Omega^{-2}$ terms. To understand scattering processes due to these operators at all energies, we would need to quantise an action with inverse powers of fields. It is not obvious how this can be done in general, and it is this issue that we address here. 

The conventional estimate of unitarity violation in the non-minimally coupled SM in the Palatini formalism assumes that the $\xi \phi^{2}/M_{Pl}^{2}$ terms in $\Omega^{2}$ can be considered small compared to 1, in which case $\Omega^{-2}$ can be Taylor expanded. Then, for example,  
\be{e9}  -\frac{1}{2 \Omega^{2}} \partial_{\mu} \phi_{1} \partial^{\mu} \phi_{1} = -\frac{1}{2} \partial_{\mu} \phi_{1} \partial^{\mu} \phi_{1} \left(1 - \frac{\xi \left(\phi_{1}^{2} + \phi_{2}^{2} \right) }{ M_{Pl}^{2}} + ...  \right)   ~.\ee
This results in a $\phi_{1} \phi_{2}$ scattering interaction\footnote{Interactions involving only $\phi_{1}$ or $\phi_{2}$ can generally be eliminated by a redefinition to a canonically normalised field.}, 
\be{e10} \frac{ \xi}{2 M_{Pl}^{2}} \partial_{\mu} \phi_{1} \partial^{\mu} \phi_{1} \phi_{2}^{2}    ~.\ee 
Dimensionally, this violates unitarity at $\Lambda \sim M_{Pl}/\sqrt{\xi}$, which is the conventional estimate of the unitarity violation scale in Palatini Higgs Inflation in the electroweak vacuum \cite{pi2}. 
However, this is based on the assumption that the field-dependent terms in $\Omega^{2}$ can be considered small compared to 1 in scattering processes at energy $E \gae M_{Pl}/\sqrt{\xi}$. We next argue that this may not be true.

The conventional analysis of tree-level $2 \rightarrow 2$ unitarity violation considers scattering of particles with exactly known energy and momentum. However, unitarity violation should also manifest itself in other scattering set-ups. For example, consider $2 \rightarrow 2$ scattering with particles scattering 'close to' a point ${\bf x}$ at energy 'close to' $E$. The incoming particles will then be described by wavepackets with energy-momentum modes close to $E$. By energy close to $E$, we mean a spread of energies $\Delta E \sim E$.  In this case the scattering cross-section would be expected to be of the same magnitude as that for particles of energy and momentum precisely equal to $E$. $\Delta E$  will correspond to a spread in the position of scattering by $\Delta x \sim 1/\Delta E \sim 1/E$.   Within the interaction volume $\Delta x^{3}$, the expectation value of the energy density will be $<\rho> \sim E/\Delta x^{3} \sim E^{4}$. The expectation value of the energy density will be related to the expectation value of the field in the interaction volume by
$<\rho> \sim E^{2} <\phi^{2}>$, therefore 
$\phi_{v} = <\phi^{2}>^{1/2} \sim E$, where $\phi_{v}$ is the expectation value of the magnitude of the field in the interaction volume. Note that we are considering the canonically normalised field in the electroweak vacuum here, which describes the incoming and outgoing particles of the scattering process occurring within the volume $\Delta x^{3}$.

Suppose we now introduce a constant background field $\overline{\phi}$. If $\overline{\phi} < \phi_{v} \sim E$, then we would not expect the background field to significantly modify the scattering process. If $\overline{\phi} > \phi_{v} \sim E$ on the other hand, we would need to consider the interaction in a background $\overline{\phi}$ with $\phi = \overline{\phi} + \phi'$, where $\phi'$, the dynamical field which describes the scattering particles, is considered to be small compared to $\overline{\phi}$ since $\phi_{v}' \sim E$.

Since the cross-sections in the two set-ups should generally be similar, it follows we should also be able to introduce a background field with $\overline{\phi} < E$ without affecting the calculation of the cross-section and amplitude in the case of exact energy-momentum states. This is the key point of our argument. In general, we can consider the action relevant to the computation of the cross-section in the two set-ups to be the same at a given scattering energy. In practice, this means that we can consider the dynamical fields to have magnitude $\phi' \sim E$ when determining the relevant expansion of the action.  

Following from this, our analysis will be based on the following assumptions: 

\vspace{0.1cm} 

\noindent (i): The scattering amplitude for $\phi$ scalars of energy $E$ is essentially unaffected by introducing a background field $\overline{\phi}$ for $\phi$ so long as $\overline{\phi} \ll E$. 

\vspace{0.1cm}

\noindent  (ii): Although the background field cannot be neglected once $\overline{\phi} \approx E$, we will assume that the magnitude of the scattering amplitude will still be correct in the limit $\overline{\phi} \approx E$. Similarly, if we consider the limit $\overline{\phi} \gg E$ and compute the amplitude, we will assume that the magnitude of the amplitude will still be correct in the limit $\overline{\phi} \approx E$. This would be expected if the scattering amplitude is continuous and varies smoothly with $E$.  

\vspace{0.1cm}

We will focus on scattering $\phi_{1} \phi_{2} \rightarrow \phi_{1} \phi_{2}$ in the electroweak vacuum at Jordan frame energies $E$ greater than the conventional Palatini unitarity violation scale $\Lambda \sim M_{Pl}/\sqrt{\xi}$.  In order to use the above assumptions, we will introduce a constant background field $\overline{\phi}_{1}$ for $\phi_{1}$ that is larger than $E$. This will allow us to expand the $\Omega^{-2}$ factor in the Einstein frame and compute the scattering amplitude. We then use Assumption (ii) to match the amplitude in the limit $\overline{\phi}_{1} \rightarrow E$ to the amplitude in the limit $\overline{\phi}_{1} \ll E$ and so obtain the scattering amplitude in the electroweak vacuum.  In effect, we are using a threshold approximation, where the calculations valid at $\overline{\phi}_{1} \ll E$ and $\overline{\phi}_{1} \gg E$ are matched at $\overline{\phi}_{1} = E \gg M_{Pl}/\sqrt{\xi}$. 

In practice, our estimate of the scattering amplitude in the electroweak vacuum when $E \gg M_{Pl}/\sqrt{\xi}$ consists of the following steps: 

\vspace{0.1cm} 

\noindent (a) Consider the process $\phi_{1} \phi_{2} \rightarrow \phi_{1} \phi_{2}$ in the electroweak vacuum with energy $E \gg M_{Pl}/\sqrt{\xi}$. 

\vspace{0.1cm} 

\noindent (b) Transform to the Einstein frame. Introduce a constant background field $\overline{\phi}_{1}$ such that $\overline{\phi}_{1} \gg E$ and derive the relevant expansion of the action. 

\vspace{0.1cm}

\noindent (c) Define canonically normalised scalars $\chi_{1}'$ and $\chi_{2}$ and a constant background field   $\overline{\chi}_{1}$ corresponding to $\overline{\phi}_{1}$. Consider scattering at Einstein frame energy $\tilde{E}$. Compute the scattering amplitude and so the energy of unitarity violation in the Einstein frame, $\tilde{\Lambda}$ .

\vspace{0.1cm} 

\noindent (d) Transform the energy to obtain the corresponding unitarity violation scale in the Jordan frame, $\Lambda$. (We will show that $\Lambda$ depends upon $\overline{\phi}_{1}$.) By Assumption (ii), this will give the correct magnitude for the unitarity violation scale in the limit $\overline{\phi}_{1} \rightarrow E$, and so the correct magnitude for the unitarity violation scale $\Lambda$ in the limit $\overline{\phi}_{1} \ll E$, corresponding to the unitarity violation scale for electroweak vacuum scattering of $\phi_{1}$ and $\phi_{2}$ scalars at $E \gg M_{Pl}/\sqrt{\xi}$.

\vspace{0.1cm}

We first derive the relevant expansion of the action.  We introduce $\overline{\phi}_{1}$ such that $\overline{\phi}_{1} \gg E \gg M_{Pl}/\sqrt{\xi}$. Since  $\phi_{1} = \overline{\phi}_{1} + \phi_{1}' \gg E$, where $\phi_{1}'$ is the dynamical part of the field and $\phi_{1}' \sim E$, we can write the Lagrangian as 
\be{x1}  - \frac{1}{2 \Omega^{2}} \partial_{\mu} \phi_{1} \partial^{\mu} \phi_{1} 
 - \frac{1}{2 \Omega^{2}} \partial_{\mu} \phi_{2} \partial^{\mu} \phi_{2}   \approx -\frac{1}{2} \frac{\partial_{\mu} \phi_{1} \partial^{\mu} \phi_{1} }{ \left(1 + \frac{\xi \phi_{1}^{2}}{M_{Pl}^{2}} \right) \left(1 + \frac{\phi_{2}^{2}}{\overline{\phi}_{1}^{2}} \right) }   -
\frac{1}{2} \frac{\partial_{\mu} \phi_{2} \partial^{\mu} \phi_{2} }{ \left(1 + \frac{\xi \phi_{1}^{2}}{M_{Pl}^{2}} \right) \left(1 + \frac{\phi_{2}^{2}}{\overline{\phi}_{1}^{2}} \right) }   ~.\ee
We transform $\phi_{1}$ to an approximately canonically normalised form $\chi_{1}$ in the usual way 
\be{x2} \frac{d \chi_{1}}{d \phi_{1}} = \frac{1}{\sqrt{1 + \frac{\xi \phi_{1}^{2}}{M_{Pl}^{2}} } }     ~.\ee
This has solution (assuming $\chi_{1} = \phi_{1}$ when $\phi_{1} = 0$) 
\be{x3} \chi_{1} = \frac{M_{Pl}}{\sqrt{\xi}} \sinh^{-1} \left( \frac{\sqrt{\xi} \phi_{1}}{M_{Pl}} \right)  ~.\ee 
The first term in the Lagrangian \eq{x1} then becomes 
\be{x5} -\frac{1}{2} \frac{ \partial_{\mu} \chi_{1} \partial^{\mu} \chi_{1} }{\left(1 + \frac{\phi_{2}^{2}}{\overline{\phi}_{1}^{2}} \right)   }    ~.\ee 
Expanding this using $\overline{\phi}_{1}^{2}  \gg  \phi_{2}^{2} \sim E^2$, we obtain the $\chi_{1}$ kinetic terms and the leading-order interaction that can lead to $\phi_{1} \phi_{2} \rightarrow \phi_{1} \phi_{2}$ scattering in the electroweak vacuum
\be{x6} -\frac{1}{2} \partial_{\mu} \chi_{1} \partial^{\mu} \chi_{1}  + \frac{1}{2} \frac{\phi_{2}^{2}}{\overline{\phi_{1}}^{2}} \partial_{\mu} \chi_{1} \partial^{\mu} \chi_{1}  +  ...    ~.\ee 
The second term of the Lagrangian \eq{x1} can be written as 
\be{x7} \frac{M_{Pl}^{2}}{2 \xi \phi_{1}^{2}} 
\frac{\partial_{\mu} \phi_{2} \partial^{\mu} \phi_{2} }
{ \left(1 + \frac{M_{Pl}^{2}}{\xi \phi_{1}^{2}} \right) \left(1 + \frac{\phi_{2}^{2}}{\overline{\phi}_{1}^{2}} \right) }  ~.\ee 
Expanding this, we obtain the $\phi_{2}$ kinetic term and leading-order interactions that can lead to $\phi_{1} \phi_{2} \rightarrow \phi_{1} \phi_{2}$ scattering 
\be{x8}  - \frac{M_{Pl}^{2}}{2 \xi \overline{\phi}_{1}^{2} } \partial_{\mu} \phi_{2} \partial^{\mu} \phi_{2} \left(1 - 2 \frac{\phi_{1}'}{\overline{\phi}_{1}}  +   3 \frac{\phi_{1}'^{2}}{\overline{\phi}_{1}^{2} } \right)  + ...   ~.\ee 
We define a canonically normalised field $\chi_{2}$ by 
\be{x9}  \chi_{2} = \frac{M_{Pl}}{\sqrt{\xi} \, \overline{\phi}_{1}} \phi_{2}  ~.\ee 
We also relate $\phi_{1}'$ to the dynamical part $\chi_{1}'$ of the $\chi_{1}$ field. In the limit where $\phi_{1}$ is large compared to $M_{Pl}/\sqrt{\xi}$, from \eq{x3} we have
\be{x10} \frac{\sqrt{\xi} \phi_{1}}{M_{Pl}} =  \sinh \left( \frac{\sqrt{\xi} \chi_{1}}{M_{Pl}} \right) \approx \frac{1}{2} e^{\frac{\sqrt{\xi} \chi_{1}}{M_{Pl}} }  ~.\ee
Therefore
\be{x11} \chi_{1} \approx \frac{M_{Pl}}{\sqrt{\xi}} \ln \left( \frac{2 \sqrt{\xi} \phi_{1}}{M_{Pl}} \right)  ~.\ee
With $\phi_{1} = \overline{\phi}_{1} + \phi_{1}'$, where $\phi_{1}' \ll \overline{\phi}_{1}$, we obtain
\be{x12} \chi_{1} \approx \frac{M_{Pl}}{\sqrt{\xi}} \ln 
\left( \frac{2 \sqrt{\xi} \overline{\phi}_{1}}{M_{Pl}} \left(1 + \frac{\phi_{1}'}{\overline{\phi}_{1}} \right) \right)  ~.\ee Therefore  
\be{x13} \chi_{1} =  
 \frac{M_{Pl}}{\sqrt{\xi}} \left( \ln 
\left( \frac{2 \sqrt{\xi} \overline{\phi}_{1}}{M_{Pl}} \right)  + \ln \left(1 + \frac{\phi_{1}'}{\overline{\phi}_{1}} \right) \right) 
\approx \frac{M_{Pl}}{\sqrt{\xi}} \ln 
\left( \frac{2 \sqrt{\xi} \overline{\phi}_{1}}{M_{Pl}} \right)  + \frac{M_{Pl}}{\sqrt{\xi}} \frac{\phi_{1}'}{\overline{\phi}_{1}}   ~.\ee 
We can then define $\chi_{1} = \overline{\chi}_{1} + \chi_{1}'$, where 
\be{x14} \overline{\chi}_{1} = \frac{M_{Pl}}{\sqrt{\xi}} \ln 
\left( \frac{2 \sqrt{\xi} \, \overline{\phi}_{1}}{M_{Pl}} \right) ~\ee 
and 
\be{x15} \chi_{1}' =  \frac{M_{Pl}}{\sqrt{\xi}} \frac{\phi_{1}'}{\overline{\phi}_{1}}    ~.\ee 
Therefore, in terms of the canonically normalised dynamical fields $\chi_{1}'$ and $\chi_{2}$, from \eq{x6} and \eq{x8} the leading order interactions leading to $\chi_{1}' \chi_{2} \rightarrow \chi_{1}' \chi_{2}$ scattering are 
\be{x16}  \frac{1}{2} \frac{\xi}{M_{Pl}^2} \chi_{2}^{2} \partial_{\mu} \chi_{1}' \partial^{\mu} \chi_{1}' + 
\frac{\sqrt{\xi}}{M_{Pl}} \chi_{1}' \partial_{\mu} \chi_{2} \partial_{\mu} \chi_{2} - \frac{3}{2} \frac{\xi}{M_{Pl}^{2}} \chi_{1}'^{2}\partial_{\mu} \chi_{2} \partial_{\mu} \chi_{2}    ~.\ee 
Dimensionally, all of these terms will lead to scattering amplitudes as a function of the Einstein frame energy $\tilde{E}$ of the form 
\be{x17} |{\cal M}| \sim \frac{\xi \tilde{E}^{2}}{M_{Pl}^{2}}   ~, \ee
either directly via the quartic interactions or from $\chi_{2}$ exchange via the cubic interaction. 
It follows that the scale of unitarity violation in the Einstein frame is given by 
\be{x18} \tilde{\Lambda} \approx \frac{M_{Pl}}{\sqrt{\xi}}    ~.\ee 
The energy in the Einstein frame $\tilde{E}$ is related to that in the Jordan frame by 
\be{x19}  \tilde{E} = \frac{E}{\Omega}   ~.\ee 
The amplitude \eq{x17} is calculated with $\overline{\phi}_{1} \gg  M_{Pl}/\sqrt{\xi}$, thus 
\be{x20}  \Omega \approx \left(1 + \frac{\xi \overline{\phi}_{1}^{2} }{M_{Pl}^{2}} \right)^{1/2} \approx \frac{ \sqrt{\xi}\, \overline{\phi}_{1}}{M_{Pl}}    ~.\ee 
Therefore the scattering amplitude in terms of the energy in the Jordan frame is\footnote{Note that this also shows that the Jordan frame unitarity violation scale in the presence of a background inflaton field $\overline{\phi}$ would be at $\Lambda \sim \overline{\phi}$.} 
\be{x21} |{\cal M}| \sim   \frac{\xi E^{2}}{\Omega^{2} M_{Pl}^{2}} = \frac{E^{2}}{\overline{\phi}_{1}^{2}}     
 ~,\ee 
and the corresponding scale of unitarity violation in the Jordan frame is 
\be{x21aa} \Lambda = \Omega \tilde{\Lambda} \sim \overline{\phi}_{1}   ~.\ee

We then apply Assumption (ii), which implies that the scattering amplitude \eq{x21} in the limit $\overline{\phi}_{1} \rightarrow E$ also gives the scattering amplitude in the limit $\overline{\phi}_{1} \rightarrow 0$, where the Jordan and Einstein frames are equivalent. Therefore the tree-level scattering amplitude for $\phi_{1}\phi_{2} \rightarrow \phi_{1} \phi_{2}$ in the electroweak vacuum at energy $E \gg M_{Pl}/\sqrt{\xi}$ is 
\be{x21a} |{\cal M}| \sim  \lim_{\overline{\phi}_{1} \to E} \, \frac{E^{2}}{\overline{\phi}_{1}^{2}} \sim 1     
 ~.\ee 
Based on this, the scattering amplitude as a function of $E$ increases as $E^{2}$ until $E \sim M_{Pl}/\sqrt{\xi}$, and then becomes a constant close to 1. 
\noindent The corresponding unitarity violation scale \eq{x21a} is\footnote{One may ask why we do not simply consider the unitarity violation scale in the Einstein frame to be the scale of unitarity violation. The reason is that we start by considering scattering of scalars at a fixed energy $E$ in the Jordan frame in the electroweak vacuum. We then introduce a background field $\overline{\phi}_{1}$,  keeping $E$ constant throughout. Each value of $\overline{\phi}_{1}$ in effect defines a different model with a different conformal transformation to the Einstein frame. In order to keep $E$ constant as $\overline{\phi}_{1}$ varies, the scattering energy in the Einstein frame must vary with $\overline{\phi}_{1}$. From the point of view of the Einstein frame, even though there is a unitarity cutoff at $\tilde{\Lambda} \sim M_{Pl}/\sqrt{\xi}$, the scattering energy in the Einstein frame varies with the background field as $\tilde{E} = E/\Omega =  E M_{Pl}/\sqrt{\xi}\, \overline{\phi}_{1} =  E \tilde{\Lambda}/\overline{\phi}_{1}$. Since $\overline{\phi}_{1} \gae E$, $\tilde{E}$ can never exceed $\tilde{\Lambda}$.}   
\be{x21b} \Lambda \sim E  ~.\ee

In general, if unitarity is violated in a theory we would expect there to be a well-defined energy of unitarity violation, with the degree of violation increasing with $E$. Therefore,  discounting a marginal unitarity violation which is constant with $E$ when $E > M_{Pl}/\sqrt{\xi}$, we interpret \eq{x21a} and \eq{x21b} to mean that in the Palatini model tree-level unitarity violation is not violated in high energy Higgs scattering in the electroweak vacuum at energies below the Planck scale. Based on this, we conclude that Palatini Higgs Inflation is likely to be a completely consistent particle theory at all energies up to the Planck scale.

\section{Conclusions}

We have considered the question of unitarity violation in the electoweak vacuum in the Palatini formulation of Higgs Inflation. Unitarity violation in scattering processes represents a fundamental breakdown of the theory, requiring that the theory is replaced by a new unitary theory before the energy of unitarity violation is reached. In Palatini Higgs Inflation, the conventional view is that unitarity in Higgs scattering in the electroweak vacuum is violated at an energy $\Lambda \sim M_{Pl}/\sqrt{\xi}$. Here we have argued that it is likely that there is no unitarity violation in Palatini Higgs Inflation in the electroweak vacuum at any scattering energy below the Planck scale. This is because the expansion of inverse terms of the conformal factor $\Omega$, which leads to the conventional estimate of unitarity violation, is not valid in scattering at the conventional energy of unitarity violation. We have estimated the scattering amplitude for Higgs scattering at energy $E \gg M_{Pl}/\sqrt{\xi}$ and find that it is at most of the order of unity.  

If true, the non-minimally coupled Standard Model in the Palatini formulation would be a completely consistent particle theory at all energies up to the Planck scale. As such, Palatini Higgs Inflation would have a significant advantage over the metric version of the theory\footnote{In \cite{gl}, it was claimed that by adding a singlet scalar to the Standard Model the unitarity problem of metric Higgs Inflation can be solved. However, as shown in the Appendix of \cite{uvjr1}, this is not the case. In the model of \cite{gl}, the SM sector is coupled to a completely separate induced gravity inflation model, with the Higgs playing no role in inflation. The resulting model is not meaningfully related to Higgs Inflation. It is important not to diminish the difficulty of the unitarity problem of metric Higgs Inflation by continuing to propagate the notion that it can be solved simply by adding a scalar particle.}
\footnote{Recently, Higgs Inflation in more a general gravity formulation, which includes both metric and Palatini as special cases, has been explored in \cite{sys1}. The argument of this paper may be relevant to determining consistent models from the range of possibilities presented by this framework.}.

Our conclusion is based on a physical argument relating the scattering amplitude of particles of exact energy $E$ to that of particles with an uncertainty $\Delta E \sim E$. As such, it does not constitute a formal proof, although we believe it strongly suggests that tree-level unitarity is conserved in Palatini Higgs Inflation up to the Planck scale. At the very least, we believe that our argument indicates that this important issue is yet to be resolved.

\end{document}